\documentclass[aps,twocolumn,prb,longbibliography,preprintnumbers,superscriptaddress,sloatfix]{revtex4-2}

\usepackage{amssymb, amsfonts,amsmath,dsfont}
\usepackage{blindtext}
\usepackage{graphicx}
\usepackage{stackengine}
\usepackage{siunitx}
\usepackage[caption=false]{subfig}
\usepackage{xcolor}
\usepackage{amsmath}
\usepackage{tabularx}
\usepackage[utf8]{inputenc}
\usepackage{tablefootnote}
\usepackage{bm}
\usepackage{color}
\usepackage{accents}
\usepackage{mathtools}
\usepackage{hyperref}
\usepackage[capitalize]{cleveref}
\usepackage{url}
\hypersetup{colorlinks=true, citecolor=blue, urlcolor=blue, linkcolor=blue}

\DeclareSIUnit{\Oe}{Oe}

\newcommand{\uvec}[1]{\mathbf{\hat{\textbf{#1} } } }
\newcommand{\op}[1]{\hat{#1}}

\newcommand\thicknm{50}

\usepackage{float}
\floatstyle{plaintop}
\restylefloat{table}

\begin{document}

\title{Anisotropy-assisted magnon condensation in ferromagnetic thin films}
\author{Therese Frostad}
\affiliation{Center for Quantum Spintronics, Department of Physics, Norwegian University of Science and Technology, NO-7491 Trondheim, Norway}
\author{Philipp Pirro}
\author{Alexander A. Serga}
\author{Burkard Hillebrands}
\affiliation{Fachbereich Physik and Landesforschungszentrum OPTIMAS, Rheinland-Pf\"{a}lzische Technische Universit\"{a}t Kaiserslautern-Landau, 67663 Kaiserslautern, Germany}
\author{Arne Brataas}
\author{Alireza Qaiumzadeh}
\affiliation{Center for Quantum Spintronics, Department of Physics, Norwegian University of Science and Technology, NO-7491 Trondheim, Norway}

\begin{abstract}
We theoretically demonstrate that adding an easy-axis magnetic anisotropy facilitates magnon condensation in thin yttrium iron garnet (YIG) films.
Dipolar interactions in a quasi-equilibrium state stabilize room-temperature magnon condensation in YIG. Even though the out-of-plane easy-axis anisotropy generally competes with the dipolar interactions, we show that adding such magnetic anisotropy may even assist the generation of the magnon condensate electrically via the spin transfer torque mechanism. We use analytical calculations and micromagnetic simulations to illustrate this effect.
Our results may explain the recent experiment on Bi-doped YIG and open a pathway toward applying current-driven  magnon condensation in quantum spintronics.
\end{abstract}

\maketitle

{\it{Introduction---.}}
Magnon condensate with nonzero momentum at room temperature is a fascinating phenomenon first observed in 2006 \cite{demokritov2006bose}.
Condensed magnons were observed at two degenerate magnon band minima of high-quality yttrium iron garnet (YIG), an easy-plane ferrimagnetic insulator with a very low magnetic dissipation
\cite{cherepanov1993saga, soumah2018ultra,PhysRevMaterials.4.024416}, as the spontaneous formation of a quasi-equilibrium and coherent magnetization dynamics in momentum space \cite{Demok-comment,PhysRevB.104.L100410,Coherent-questions,Troncoso_2012}.
To generate magnon condensate, nonequilibrium magnons must be pumped into the system by an \emph{incoherent} stimulus such as parametric pumping
\cite{anderson1955instability,suhl1956nonlinear, demokritov2006bose, demidov2008magnon, rezende2009theoryco, nowik2012spatially, serga2014bose, bozhko2016supercurrent, sun2016unconventional, dzyapko2016high, borisenko2020direct}, rapid cooling of thermal magnons \cite{Schneider,PhysRevB.104.L140405,Safranski}, and spin-transfer torque
\cite{divinskiy2021evidence,PhysRevLett.127.237203,PhysRevLett.131.156701,
demidov2011control,  demidov2017magnetization,bender2012electronic,bender2014dynamic,tserkovnyak2016bose}.
Above a critical nonequilibrium magnon density, magnons may finally (quasi)thermalize to form a quasi-equilibrium magnon condensate at the bottom of magnon bands.

The study of magnon condensation is not only interesting from an academic point of view but also of great importance in various areas of emerging quantum technology and applied spintronics \cite{PhysRevB.90.144419, andrianov2014magnon, mohseni2022classical, bozhko2016supercurrent,Bunkov1}. Therefore, it is crucial to clarify the intricate microscopic mechanisms at play and to present theoretical proposals to electrically control the generation of magnon condensate.

At low magnon densities, the interaction between magnons is weak, and they behave as free quasiparticles. But when the magnon population increases, as in magnon condensation experiments, the interactions between magnons become stronger and more crucial. Moreover, nonlinear magnon interactions facilitate the quasi-thermalization process of injected nonequilibrium magnons. A stable and steady quasi-equilibrium magnon condensation requires an effective repulsive interaction between injected magnon quasiparticles. It is known that in a system mainly influenced by Heisenberg exchange interactions, interaction between magnons is attractive. However, it was shown that dipolar interactions may assist in the generation of a
\emph{metastable} double-degenerate magnon condensate in YIG
\cite{PhysRevB.96.064438,demokritov2008quantum, demidov2008observation,rezende2009theoryco,rezende2009theorymi,nowik2012spatially,serga2014bose,tupitsyn2008stability,li2013phase,salman2017microscopic,hick2010bose,demidov2007thermalization}.

Recently, it was theoretically shown that the quasi-thermalization time of magnon condensation is reduced in confined nanoscopic systems \cite{mohseni2020bose}.
It was also demonstrated that the lateral confinement in YIG thin films enhances the dipolar interaction along the propagation direction of magnons and causes a deeper band depth, i.e., the difference between the ferromagnetic resonance (FMR) and magnon band minima. Increasing the lifetime of the magnon condensate was attributed to this enhancement of the band depth \cite{mohseni2020bose}.

In another recent achievement in magnon condensation experiments, Divinsky et al. \cite{divinskiy2021evidence} found evidence of condensation of magnons by spin-transfer torque mechanism. They introduced a small perpendicular magnetocrystalline anisotropy (PMA) through bismuth doping in the thin film of YIG, while the magnetic ground state still resides within the plane. This discovery opens a route toward electrical control of magnon condensation.

However, the interplay between the dipolar interactions, which was previously shown to be essential for the stability and thermalization of magnon condensation, and the counteracting out-of-plane easy-axis magnetic anisotropy is so far uncharted.
In this Letter, we analyze the stability of condensate magnons in the presence of a PMA in YIG. We present simulations within the Landau-Lifshitz-Gilbert framework \cite{lakshmanan2011fascinating, landau1992theory,gilbert1955lagrangian} that support our analytical calculations.

{\it{Model---.}}
We consider a thin ferromagnetic film in the $y-z$ plane to model YIG. The magnetic moments are directed along the $z$ direction by an in-plane external magnetic field of strength $H_0$. The magnetic potential energy of the film contains contributions from the isotropic exchange interaction $\mathcal{H}_\text{ex}$,
Zeeman interaction $\mathcal{H}_\text{Z}$, dipolar interaction $\mathcal{H}_\text{dip}$, and additionally a PMA energy $\mathcal{H}_\text{an}$ in the $x$ direction, normal to the film plane. YIG has a weak in-plane easy-axis that can be neglected compared to the other energy scales in the system.
The total spin Hamiltonian of the system reads,
\begin{equation}
\mathcal{H} = \mathcal{H}_\text{ex} + \mathcal{H}_\text{Z} + \mathcal{H}_\text{dip} + \mathcal{H}_\text{an}.
\end{equation}
The PMA energy is given by,
\begin{equation}
\mathcal{H}_\text{an} = -K_\text{an} \sum_{j} (\bm{S}_j \cdot \hat{x})^2,
\end{equation}
where $K_\text{an}>0$ is the easy-axis energy, $\hbar \bm{S}_j$ is the vector of spin operator at site $j$, and $\hbar$ is the reduced Planck constant. Details of the Hamiltonian can be found in the Supplemental Material (SM) \cite{SM}.

The Holstein-Primakoff spin-boson transformation \cite{holstein1940field} allows us to express the spin Hamiltonian in terms of magnon creation and annihilation operators. The amplitude of the effective spin per unit cell in YIG at room temperature is large $S \approx 14.3 \gg 1$,
\cite{li2013phase, streib2019magnon,maier2017temperature}, and thus we can expand the spin Hamiltonian in the inverse powers of the spin $S$, which is equivalent to the semiclassical regime.
Up to the lowest order in nonlinear terms, the magnon Hamiltonian $\mathcal{H}$ of a YIG thin film can be expressed as the sum of two components:
$\mathcal{H}_2$ and $\mathcal{H}_4$.
The former represents a noninteracting magnon gas comprising quadratic magnon operators. The latter, on the other hand, constitutes nonlinear magnon interactions characterized by quartic magnon operators; see SM for details \cite{SM}. Note that three-magnon interactions are forbidden in our geometry by the conservation laws \cite{PhysRevB.38.11444}

{\it{Magnon dispersion of YIG with a finite PMA---.}}
The magnon dispersion in YIG is well known and has been studied extensively in both experimental and theoretical works \cite{princep2017full, serga2010yig, cherepanov1993saga}.
Magnons traveling in the direction of the external magnetic field have the lowest energy. These so-called backward volume magnetostatic (BVM) magnons have a dispersion with double degenerate minima at finite wavevectors $q_z = \pm Q$.
When pumping magnons into the thin film, the magnons may thermalize and eventually form a condensate state in these two degenerate minima with opposite wavevectors.

The noninteracting magnon Hamiltonian and the dispersion of BVM magnons, along the $z$ direction, in the presence of a finite PMA reads,
\begin{subequations}
\begin{align}
\mathcal{H}_2 &= \sum_{q_z} \hbar \omega_{q_z} \op{c}_{q_z}^\dagger \op{c}_{q_z}, \\
\hbar \omega_{q_z} &= \sqrt{A_{q_z}^2 - B_{q_z}^2},
\label{eq:disprel}
\end{align}
\end{subequations}
where $\op{c}_{q_z}^\dagger (\op{c}_{q_z})$ are  magnon creation (annihilation) operators, which are Bogoliubov bosons \cite{SM}, and
\begin{subequations}
\begin{align}
A_{q_z} & = D_\text{ex} q_z^2 + \gamma (H_0 + 2 \pi M_S f_q) - K_\text{an} S, \\
B_{q_z} & = 2 \pi M_S f_q - K_\text{an} S.
\label{eq:AkBk}
\end{align}
\end{subequations}
Here, $D_{\rm{ex}}=J_{\rm{ex}}Sa^2$ is the exchange constant, $J_{\rm{ex}}$ is the Heisenberg exchange coupling and $M_S = \gamma \hbar S / a^3$ is the saturation magnetization, where $\gamma = \SI{1.2e-5}{\eV \per \Oe}$ is the gyromagnetic ratio, and $a=12.376${\AA} is the lattice constant of YIG.
The form factor $f_q = (1-e^{-|q_z| L_x})/(|q_z| L_x)$ stems from dipolar interactions in a thin magnetic film with thickness $L_x$ \cite{buijnsters2018two, zabel2012magnetic}.

\begin{figure}
\includegraphics[width = 1.0\columnwidth]{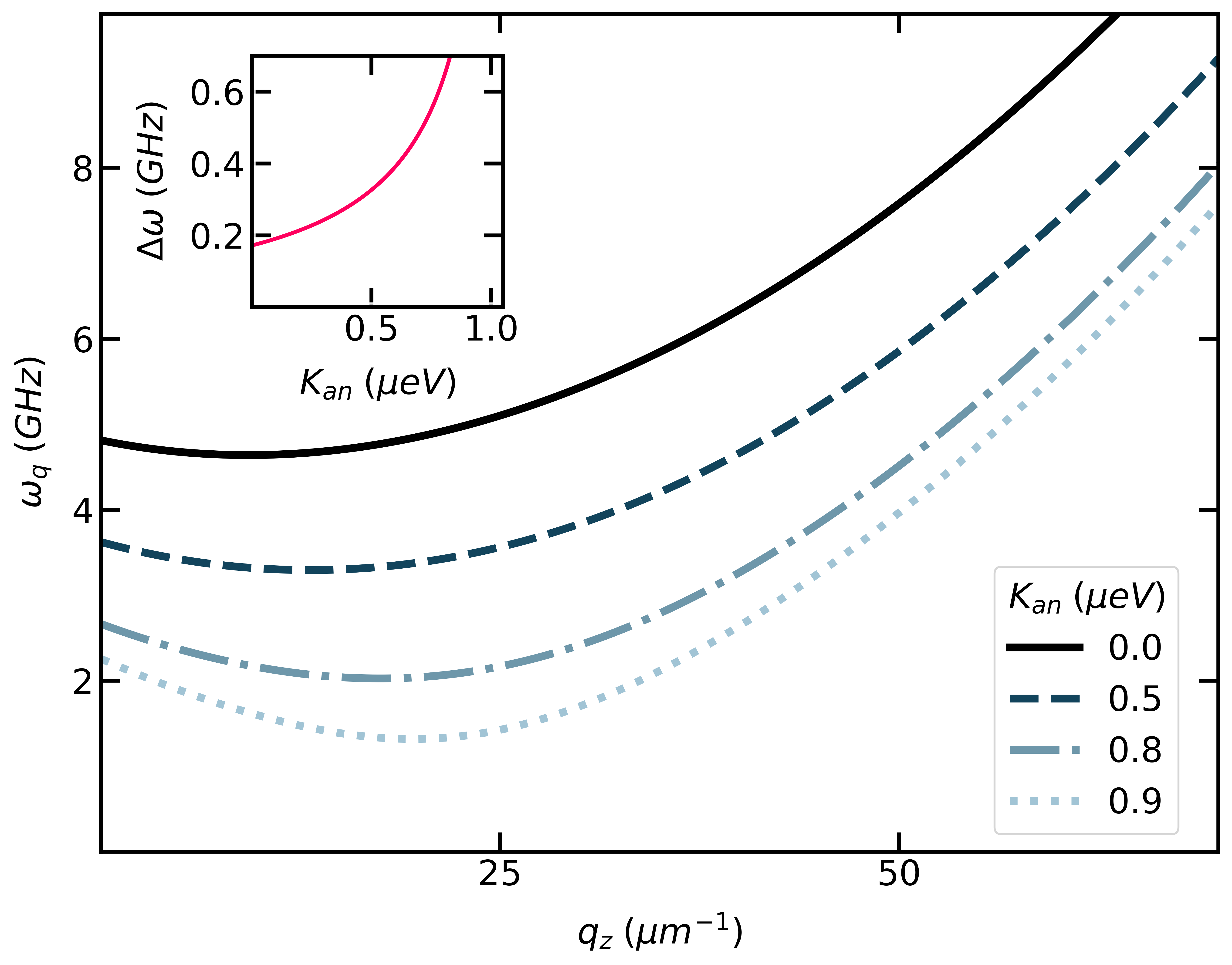}
\caption{
The analytical dispersion of noninteracting BVM magnons in a YIG thin film for various PMA strengths, Eq. (\ref{eq:disprel}). The inset shows the depth of the magnon band minima as a function of the PMA strength.
We set the thickness $L_x = \SI{\thicknm}{\nano \meter}$ and the magnetic field in the z direction $H_0 = \SI{1}{\kilo \Oe}$.
} \label{fig:disprels}
\end{figure}

Figure \ref{fig:disprels} shows the effect of PMA on the magnon dispersion of YIG. PMA decreases the FMR frequency $\omega_{{q_z}=0}$, in addition to a more significant decrease in the magnon band minima $\omega_{{q_z}= \pm Q}$.
Therefore the band depth $\Delta \omega=\omega_{{q_z}=0} - \omega_{{q_z}=\pm Q}$ is increased.
The position of the band minima at $q_z = \pm Q $ is also shifted to larger momenta. In addition, the curvature of the minima increases as a function of the anisotropy strength. Above a critical PMA, $K^{c_2}_{\text{an}}$, the magnetic ground state is destabilized and the in-plane magnetic state becomes out-of-plane. We are interested in the regime in which the magnetic ground state remains in the plane, and thus the effective saturation magnetization is positive
$M_{\rm{eff}} = M_S - 2K_\text{an}/(\mu_0 M_S) >0$.

The effect of PMA on magnon dispersion resembles the effect of confinement in the magnon spectra of YIG. In Ref. \onlinecite{mohseni2020bose}, it was shown that transverse confinement in a YIG thin film leads to an increase of the FMR frequency, the band depth, as well as shifting the band minima to higher momenta while the magnon band gap at the band minima is also increased. It was shown that this change of the spectrum in confined systems increases the magnon condensate lifetime. Therefore, we expect PMA to increase the magnon condensate lifetime and assist in the generation of magnon condensation.

{\it{Nonlinear magnon interactions in the presence of PMA---.}}
To check the stability of condensate magnons at low energy, we should show that the interactions do not destabilize and destroy magnon condensation.
To achieve this goal, we turn on the magnon interaction between the condensate magnons at $q_z = \pm Q$.
This magnon interaction consists of intra- and inter-band contributions,
$\mathcal{H}_4 = \mathcal{H}_4 ^{\text{intra}} + \mathcal{H}_4 ^{\text{inter}}$,
where
\begin{subequations} \label{inter-intra}
\begin{align}
&\mathcal{H}_4 ^{\text{intra}}=
A (
\op{c}_{Q}^\dagger \op{c}_{Q}^\dagger \op{c}_{Q} \op{c}_{Q} +
\op{c}_{-Q}^\dagger \op{c}_{-Q}^\dagger \op{c}_{-Q} \op{c}_{-Q}
), \\
&\mathcal{H}_4 ^{\text{inter}}=
2 B (
\op{c}_{Q}^\dagger \op{c}_{-Q}^\dagger \op{c}_{Q} \op{c}_{-Q} )
+ C (
\op{c}_{Q}^\dagger \op{c}_{-Q} \op{c}_{Q} \op{c}_{-Q} \nonumber \\
&
\op{c}_{-Q}^\dagger \op{c}_{-Q} \op{c}_{-Q} \op{c}_{Q} + \text{H.c.}
)
+D (
\op{c}_{Q}^\dagger \op{c}_{Q}^\dagger \op{c}_{-Q}^\dagger \op{c}_{-Q}^\dagger
+ \text{H.c.}
).
\label{eq:H4m}
\end{align}
\end{subequations}
The intraband magnon interaction, parametrized by $A$, preserves magnon number. However, the interband magnon interaction includes both a magnon conserving contribution, parametrized by $B$, and nonconserving contributions, parametrized by $C$ and $D$, see SM \cite{SM}.
The interaction amplitudes are given by
\begin{subequations} \label{eq:prefactorsD}
\begin{align}
A = &
-\frac{\gamma \pi M_S}{SN}
\big[(\alpha_1+\alpha_3)f_Q - 2\alpha_2(1-f_{2Q}) \big] \nonumber \\
& -\frac{D_{\rm{ex}}Q^2}{2SN}(\alpha_1-4\alpha_2)
+\frac{K_{\rm{an}}}{2N}(\alpha_1+\alpha_3) \label{eq:prefactorsA}, \\
B = & \frac{\gamma 2 \pi M_S}{SN}
\big[(\alpha_1-\alpha_2)(1-f_{2Q}) - (\alpha_1-\alpha_3)f_Q) \big] \nonumber \\
& +\frac{D_{\rm{ex}}Q^2}{2SN}(\alpha_1-2\alpha_2)
+\frac{K_{\rm{an}}}{N}(\alpha_1+\alpha_3) \label{eq:prefactorsB}, \\
C = & \frac{\gamma \pi M_S}{2SN}
\big[(3\alpha_1+3\alpha_2 +4\alpha_3)f_{Q} - \frac{8}{3}\alpha_3 (1-f_{2Q}) \big] \nonumber \\
& +\frac{D_{\rm{ex}}Q^2}{3SN} \alpha_3
+\frac{K_{\rm{an}}}{4N}(3\alpha_1+3\alpha_2 + 4\alpha_3) \label{eq:prefactorsC}, \\
D = & \frac{\gamma \pi M_S}{2SN}
\big[(3\alpha_1+3\alpha_2 +4\alpha_3)f_{Q} - 2\alpha_2 (1-f_{2Q}) \big] \nonumber \\
& +\frac{D_{\rm{ex}}Q^2}{2SN} \alpha_2
+\frac{K_{\rm{an}}}{2N}(3\alpha_2+\alpha_3). 
\end{align}
\end{subequations}
Here, $N$ is the total number of spin sites.
The dimensionless parameters $\alpha_1$, $\alpha_2$, and $\alpha_3$ are related to the Bogoliubov transformation coefficients, listed in SM \cite{SM}.

An off-diagonal long-rage order characterizes the condensation state.
The condensate state is a macroscopic occupation of the ground state and can be represented by a classical complex field. Therefore, to analyze the stability of the magnon condensate, we perform Madelung's transform $\op{c}_{\pm Q} \rightarrow \sqrt{N_{\pm Q}}e^{i\phi_{\pm Q}}$,
in which the macroscopic condensate magnon state is described with a coherent phase $\phi_{\pm Q} $ and a population number $N_{\pm Q}$
\cite{li2013phase, salman2017microscopic}.
The total number of condensed magnons is $N_c = N_{+Q} + N_{-Q}$,
while the distribution difference is $\delta =  N_{+Q} - N_{-Q}$. $N_c$ is set in the system by an external magnon pumping mechanism and is a constant.
We also define the total phase as $\Phi = \phi_{+Q} + \phi_{-Q}$. \\
Finally, the macroscopic four-magnon interaction energy of condensed magnons is expressed as ,
\begin{align}
\mathcal{V}_4(\delta,\Phi) =& \frac{N_c^2}{2} \big[
A+B+2C \cos{\Phi} \sqrt{1-\frac{\delta^2}{N_c^2} } \nonumber \\
& + D \cos{2 \Phi} -  \big(B-A+D \cos{2 \Phi} \big) \frac{\delta^2}{N_c^2}
\big].
\label{eq:H4mad}
\end{align}
This expression is similar to the one recently obtained without PMA \cite{Pokrovsky-new}, but the interaction amplitudes, Eq. (\ref{eq:prefactorsD}), depend on the PMA through the Bogoliubov coefficients, see SM \cite{SM} .

\begin{figure}
\includegraphics[width=0.8\columnwidth]{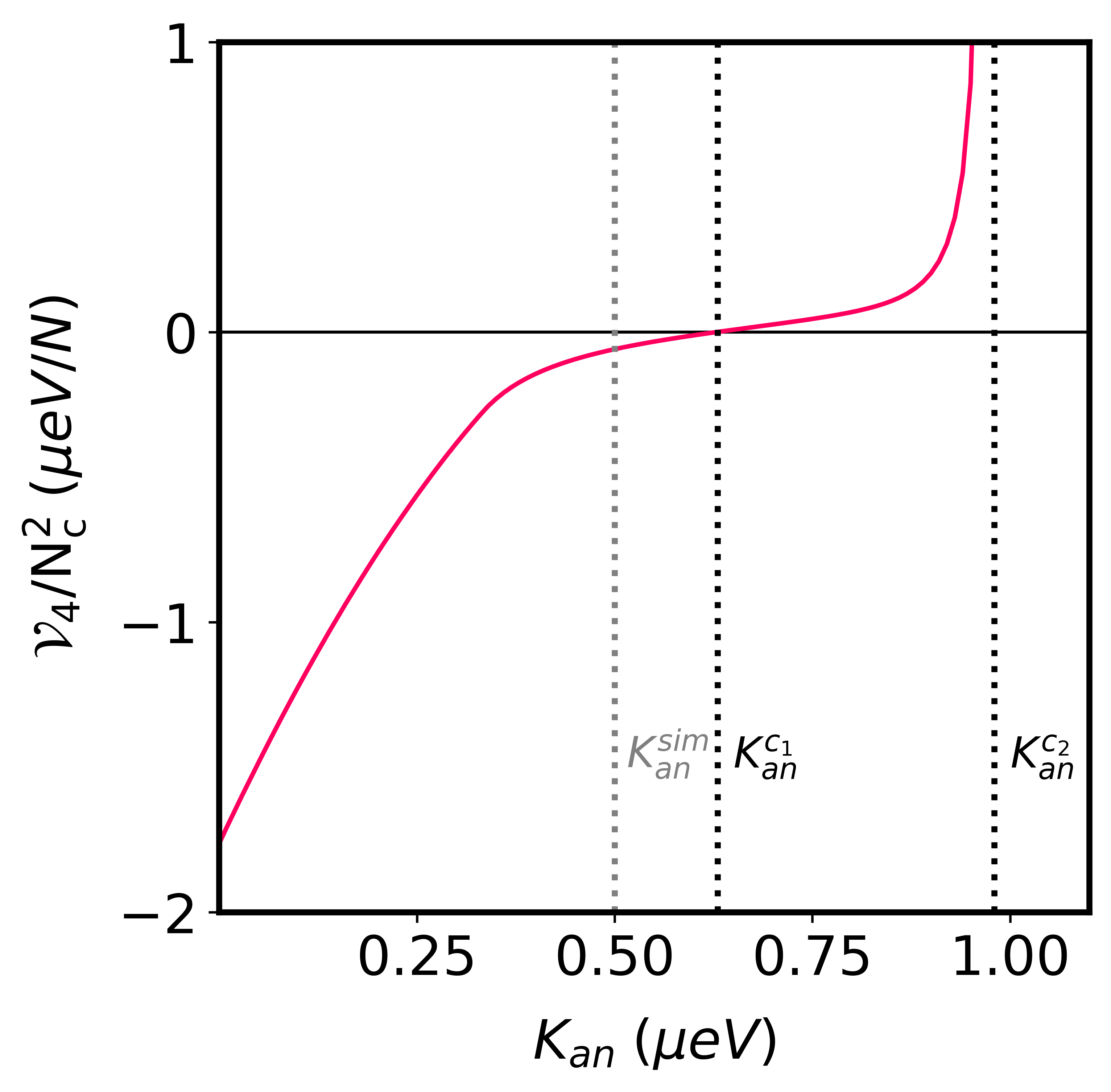}
\centering
\caption{
The analytical nonlinear interaction energy of magnon condensate state, \cref{eq:H4mad}, as a function of the PMA strength. $N$ and $N_c$ are the total spins and condensate magnons, respectively.
$K^{c_1}_{\rm{an}}$ represents the critical value of the PMA at which the sign of nonlinear interaction energy is changed.
On the other hand,
$K^{c_2}_{\rm{an}}$ corresponds to the critical value of PMA at which the in-plane magnetic ground state becomes unstable.
We set $L_x = \SI{\thicknm}{\nano \meter}$ and $H_0 = \SI{1}{\kilo \Oe}$.
$K^{sim}_{\rm{an}} = \SI{0.5}{\mu \eV}$ denotes the PMA used in our micromagnetic simulations.
} \label{fig:plot1d}
\end{figure}

We now look at the total interaction energy and amplitudes of condensate magnons in more detail.
Figure \ref{fig:plot1d} shows the effective interaction potential of condensate magnons as a function of the PMA.
In a critical PMA strength,
$K^{c_1}_\text{an}$, the sign of the interaction changes. This means that below $K^{c_1}_\text{an}$, the interaction reduces the total energy of condensate magnons while above $K^{c_1}_\text{an}$ the interaction increases its energy.
This critical anisotropy is well below the critical magnetic anisotropy strength $K^{c_2}_\text{an}$ that destabilizes the in-plane magnetic ground state.
In the following, we consider a PMA strength below the critical anisotropy $K_{\text{an}} < K^{c_1}_\text{an}$.

The interacting potential energy of the condensate magnons, Eq. (\ref{eq:H4mad}), has five extrema, $\partial \mathcal{V}_4(\delta_i,\Phi_i)=0$, at,
\begin{subequations}
\label{eq:minimas}
\begin{align}
& \delta_1 = 0, \Phi_1 = 0; \\
& \delta_2 = 0, \Phi_2 = \pi; \\
& \delta_3 = 0, \Phi_3 = \cos^{-1}(-\frac{C}{D}); \\
& \delta_4 = N_c \big[ 1 -(\frac{C}{B-A+D})^2  \big]^{\frac{1}{2}}, \Phi_4 = 0; \\
& \delta_5 = \delta_4, \Phi_5 = \pi .
\end{align}
\end{subequations}
$\delta_i=0$ indicates condensate states with symmetric magnon populations in the two magnon band minima while $\delta_i\neq 0$ represents states with nonsymmetrical magnon populations.
Whether any of these extrema represents the actual minimum of the interacting potential energy, i.e., $\partial^2 \mathcal{V}_4(\delta_i,\Phi_i)>0$, depends on the system parameters. Finding these minima allows us to construct the phase diagram for magnon condensate.

\begin{table}
\begin{tabularx}{\columnwidth}{lllll}
\hline \hline
Parameter & Symbol & Value \\
\hline
Saturation magnetization & $4 \pi M_S $ & $\SI{1.75}{\kilo \Oe}$ \\
Effective spin & $S$ & 14.3 \\
Exchange constant & $D_{\rm ex}$ & $\SI{0.64}{\eV \AA \squared}$ \\
Gilbert damping parameter  & $\alpha$ & $10^{-3}$ \\
\hline \hline
\end{tabularx}
\caption{The material parameters used in the micromagnetic simulations.}
\label{tab:paramMat}
\end{table}

\begin{figure}
\subfloat[]
{\includegraphics[width=0.9\columnwidth]{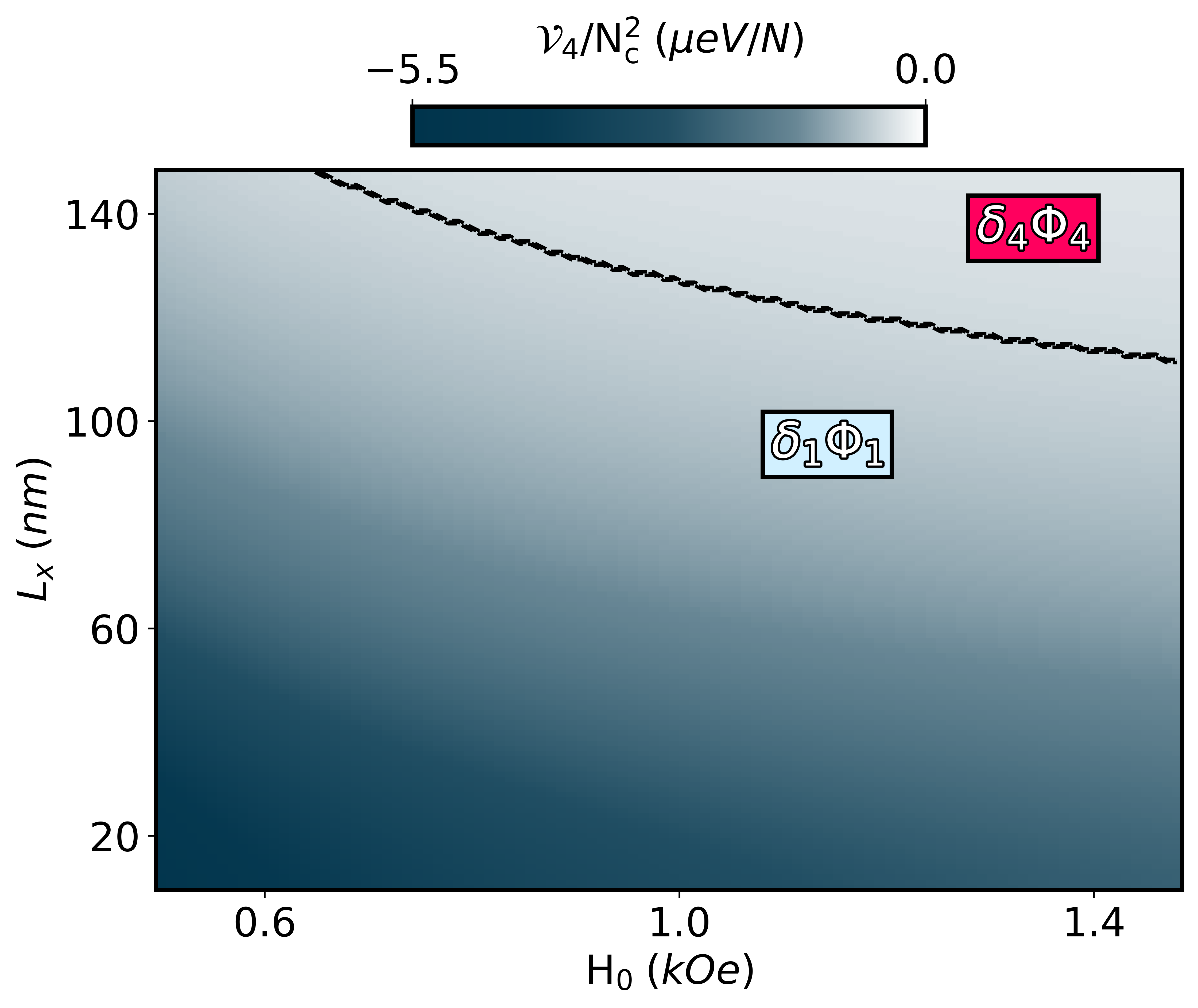}}
\hfill
\subfloat[]
{\includegraphics[width=0.9\columnwidth]{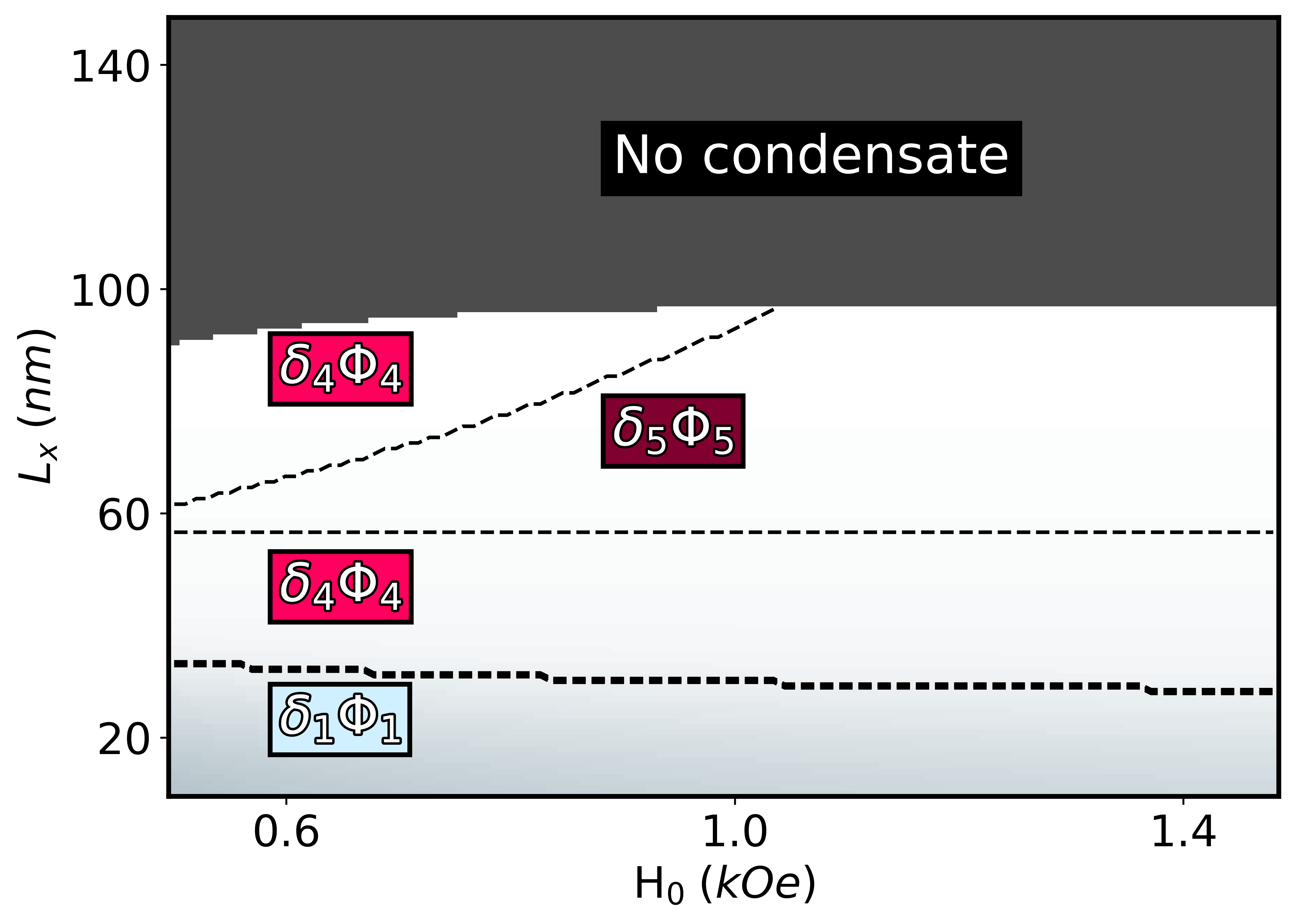}}
\caption{
The theoretical phase diagram of the condensate magnons in the absence (a) and presence (b) of PMA.
We plot the magnon interaction energy $\mathcal{V}_4/N_c^2$,
\cref{eq:H4mad}, as a function of the film thickness $L_x$ and the magnetic field strength $H_0$, applied along the $z$ direction. Different states are labeled based on different extrema listed in Eq. (\ref{eq:minimas}).
The dashed black lines indicate the boundaries between the different condensate phases, \cref{eq:minimas}. We set $K_{\rm{an}} = \SI{0.5}{\micro \eV}$ in (b).
}
\label{fig:H4mDiag}
\end{figure}

{\it{Phase diagram for magnon condensate---.}}
Now, we theoretically explore the (meta)stability of the magnon condensate as a function of the film thickness $L_x$ and the magnetic field strength $H_0$, applied along the $z$ direction in the plane, using the YIG spin parameters, see \cref{tab:paramMat}. We characterize a (meta)stable state of a magnon condensate in the phase diagram as one that minimizes the interaction potential $\mathcal{V}_4$ while satisfying the condition $\mathcal{V}_4<0$, ensuring a reduction in the total magnon condensate energy at $q_z=\pm Q$ through interactions.

First, we present the phase diagram for magnon condensation in YIG, in the absence of PMA, in \cref{fig:H4mDiag}a. The thinner films are expected to have a symmetric distribution of magnons between the two magnon band minima, the state with $\delta_1=0$, and only thicker films with larger applied magnetic fields tend to have nonsymmetric magnon populations, the state with $\delta_4 \neq 0$. This phase diagram is in agreement with previous studies
\cite{li2013phase, Pokrovsky-new}.

Next, we add a PMA, with strength $K_{\rm an} = \SI{0.5}{\mu \eV}$, and plot the phase diagram of the magnon condensate in \cref{fig:H4mDiag}b for different thicknesses. Compared to the case without PMA, we see that the condensate magnons can only be stabilized for thinner films, and within our material parameters, we do not have any metastable condensation above $90$nm since the sign of the total interaction energy becomes positive.
In addition, we have a richer phase diagram in the presence of PMA.
PMA tends to push the magnon condensate within our material parameters toward a more nonsymmetric population distribution between the two magnon band minima, states with $\delta_4 \neq 0$ and $\delta_5 \neq 0$. Since both minima are degenerate, there is an oscillation of magnon population between these two minima.
In very thin films, less than 30 nm, we may have a symmetric condensate magnon state, $\delta_1=0$, in our system.

This phase diagram shows that in the presence of a PMA, magnon condensate can be still survived as a metastable state. In addition, as we discussed earlier, a PMA increases the band depth and reduces the curvature of
noninteracting magnon dispersion, see \cref{fig:disprels}, leading to an enhancement of the condensate magnon lifetime.
Thus, we expect that introducing a small PMA into a thin film of YIG facilitates the magnon condensation process. It is worth mentioning that the stabilizing condensated magnons in thinner films with finite PMA, is not a real problematic issue since the injection of magnons into the system by electrical means is more efficient in thin films.

\begin{figure}
\subfloat{
\def\stackalignment{l}
\topinset{\bfseries(a)}{
\includegraphics[width=0.45\columnwidth]{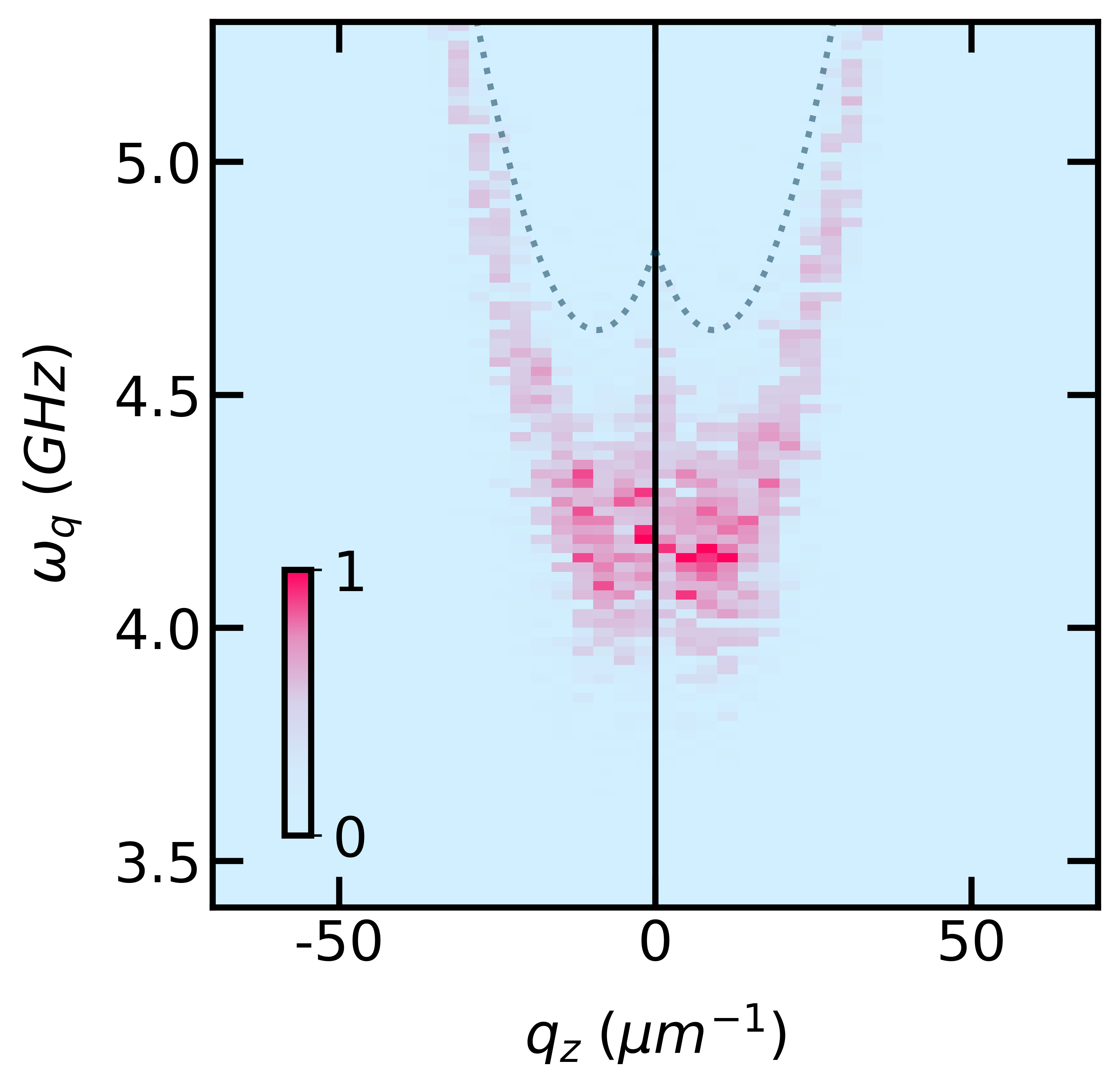}
}{0.1in}{0.1in}
}
\subfloat{
\def\stackalignment{l}
\topinset{\bfseries(b)}{
\includegraphics[width=0.45\columnwidth]{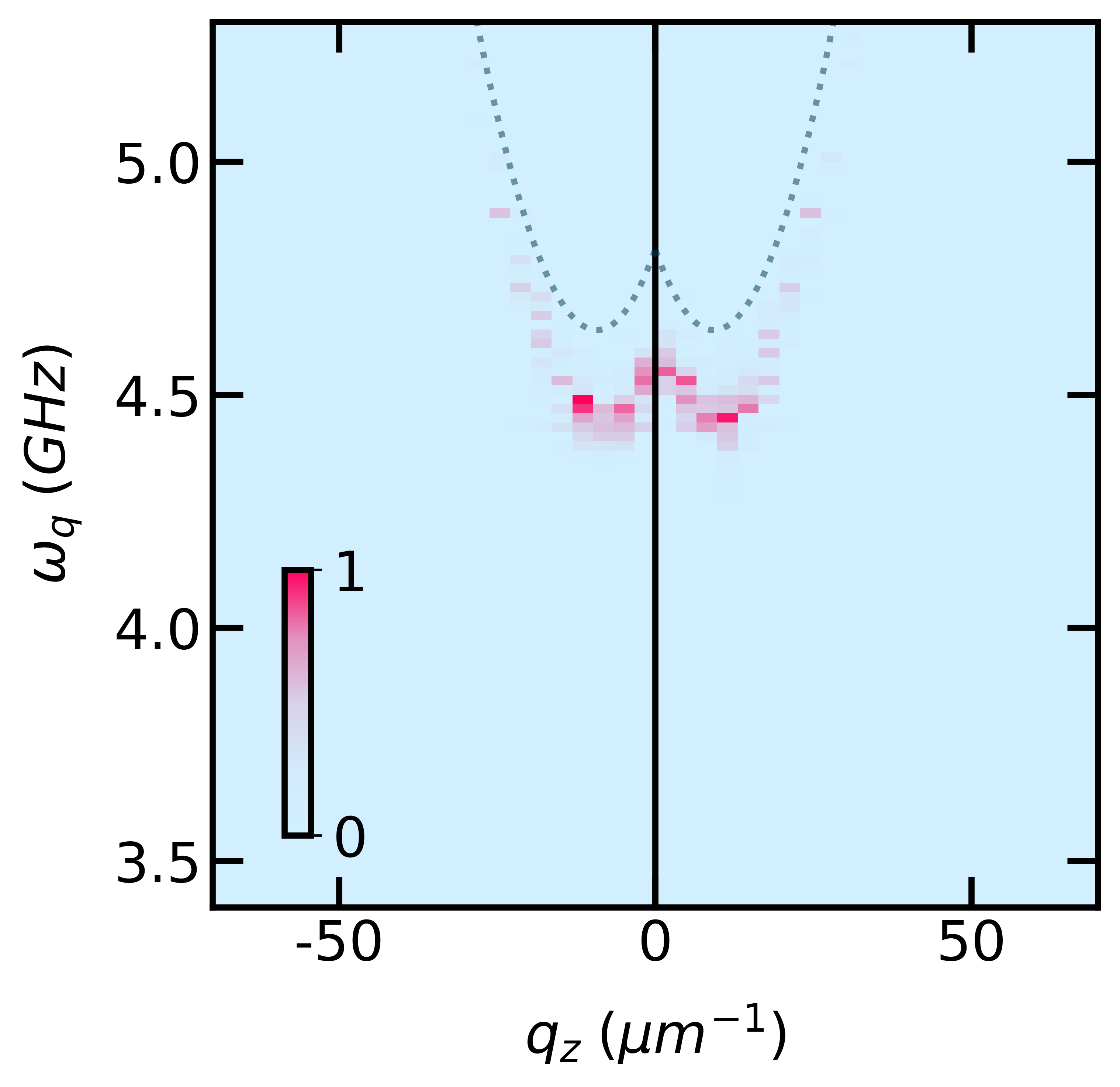}
}{0.1in}{0.1in}
}
\\
\subfloat{
\def\stackalignment{l}
\topinset{\bfseries(c)}{
\includegraphics[width=0.45\columnwidth]{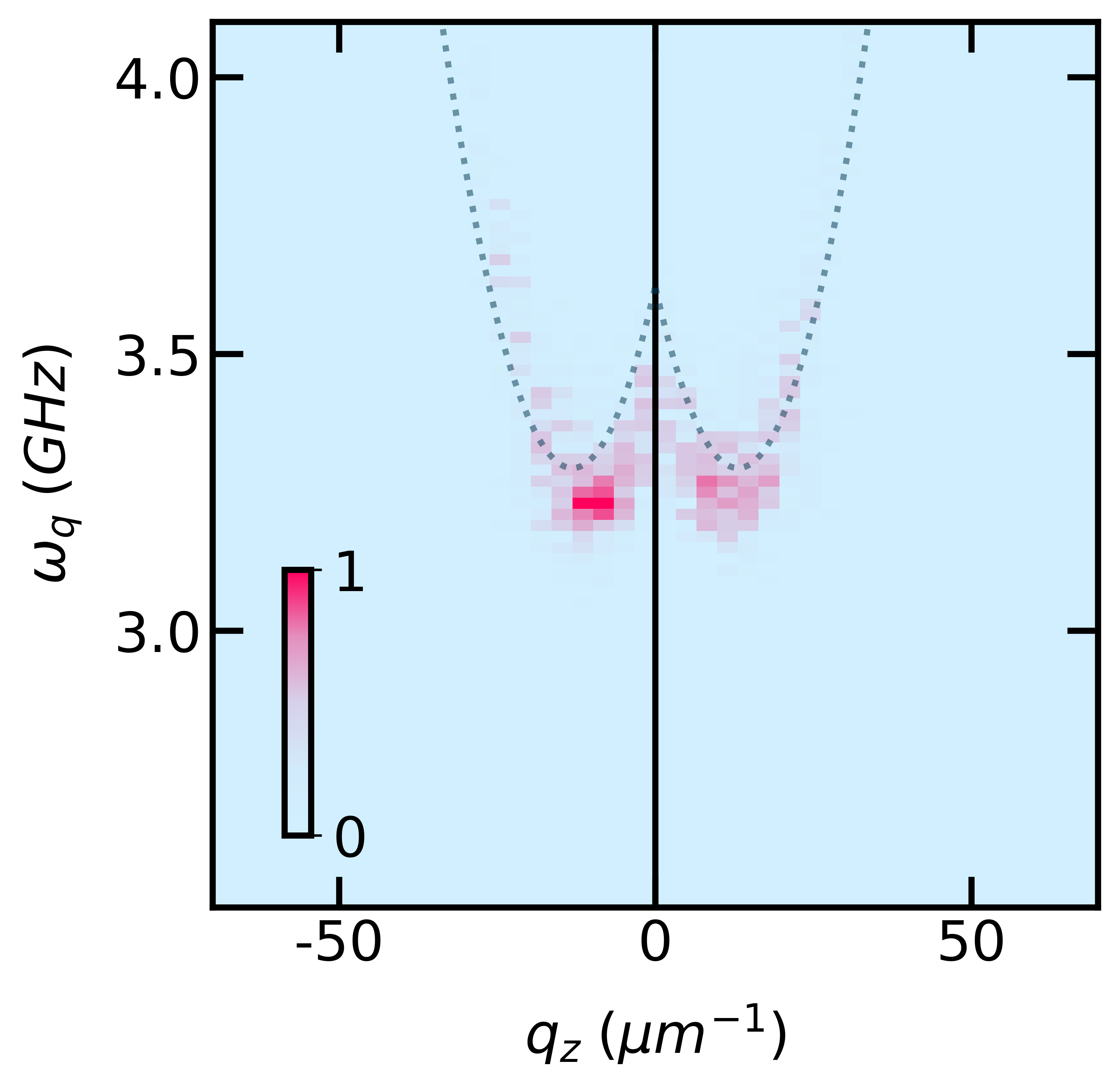}
}{0.1in}{0.1in}
}
\subfloat{
\def\stackalignment{l}
\topinset{\bfseries(d)}{
\includegraphics[width=0.45\columnwidth]{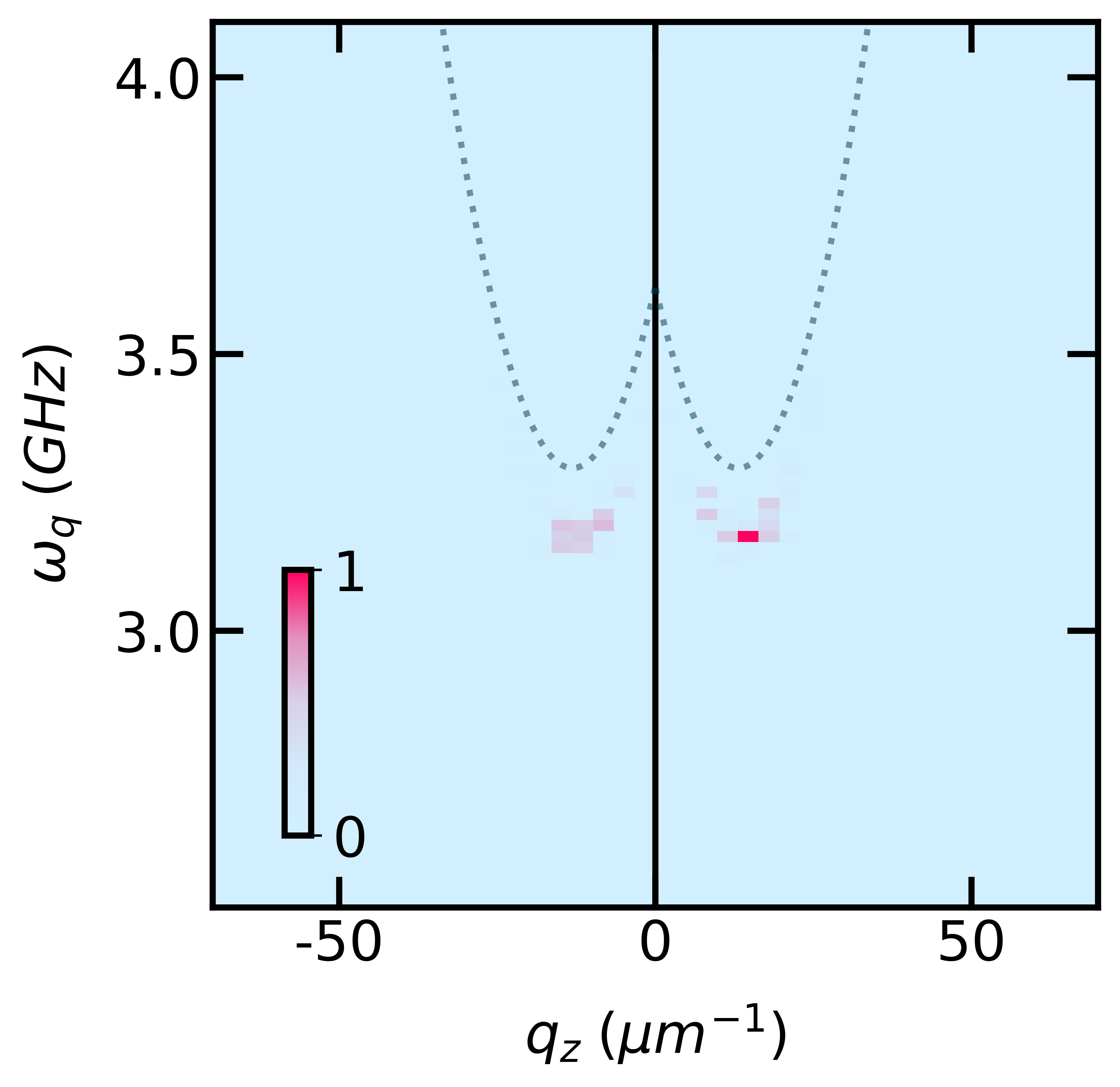}
}{0.1in}{0.1in}
}
\caption{
Micromagnetic simulation of nonequilibrium magnon distribution injected by spin torque mechanism for a YIG thin film with a thickness of $L_x=\SI{\thicknm}{\nano\meter}$ and lateral sizes of $L_y=L_z=\SI{5}{\micro\meter}$,in the presence of an external magnetic field along the $z$ direction $H_0=\SI{1}{\kilo \Oe}$.
(a) and (b) show magnon distributions of the initial nonequilibrium injected magnons and final quasi-equilibrium magnon condensate steady state, respectively, when $K_\text{an} = 0$.
(c) and (d) show magnon distributions of initial nonequilibrium excited magnons and final quasi-equilibrium magnon condensate steady state, respectively, when $K_\text{an} = \SI{0.5}{\micro \eV}$.
The dotted line indicates the analytical dispersion relation of noninteracting magnons, Eq. \ref{eq:disprel}. Because of magnon-magnon interactions, the simulated magnon dispersion has a nonlinear spectral shift compared to the analytical noninteracting magnon dispersion.
Although the duration of magnon pumping by spin-transfer torque is the same in the absence or presence of the PMA, the critical torque amplitude is lower in the presence of PMA.
}
\label{fig:fourier}
\end{figure}

{\it{Micromagnetic simulation of magnon condensate---.}}
To validate our theoretical predictions and illustrate the facilitation of magnon condensate formation by incorporating a PMA, we conducted a series of micromagnetic simulations. Simulations were performed using MuMax$^3$ \cite{vansteenkiste2014design}, which solves the semiclassical Landau-Lifshitz-Gilbert (LLG) equation that describes magnetization's precessional motion; see SM \cite{SM}. In the limit of large $S$, it is important to note that we enter the semiclassical regime and hence the LLG equation may effectively capture and accurately  describe the spin dynamics of the system.
In our ferromagnetic thin film simulation, magnons are excited via spin-transfer torque at zero temperature, eliminating thermal magnons. The nonequilibrium magnons in the film are the result of injection of spin current across the sample surface \cite{divinskiy2021evidence}. Optimal spin torque strength ensures that the magnon population reaches the critical density required for forming condensed magnons; see SM for simulation details \cite{SM}.
By introducing spin torque into the system, we excite magnons with varying wavevectors and frequencies, as illustrated in Figs. \ref{fig:fourier}(a) and \ref{fig:fourier}(c). A portion of these nonequilibrium magnons undergoes thermalization through nonlinear magnon-magnon interactions, leading to the establishment of a stable and quasi-equilibrium state of condensed magnons located at the minima of the magnon band spectra, $\pm Q$, as depicted in Figs. \ref{fig:fourier}(b) and \ref{fig:fourier}(d). This sharp peak in the number of magnons at the band minima is a signature of magnon condensate.

The numerical simulations confirm the supportive role of PMA in the condensation process.
First, there is a reduction in the threshold of spin-transfer torque necessary to inject the critical magnon density into the system, enabling the system to attain said critical magnon density even at lower torque amplitudes.
Second, the final condensate magnons in the presence of the PMA are more localized around the band minima than in the absence of PMA.
Simulations also indicate that PMA shifts the population of condensate magnons from a symmetric distribution between two band minima to a nonsymmetric distribution, Fig. \ref{fig:fourier}.
This agrees with the analytical phase diagram in Fig. \ref{fig:H4mDiag}(b).

{\it{Summary and concluding remarks---.}}
Dipolar interactions are assumed to be relevant to stabilizing the magnon condensate within YIG. The presence of a PMA is expected to counteract dipolar interactions. In this Letter, we show that even at intermediate strengths of the PMA field, a magnon condensate state can exist as a metastable state.
We note that the anisotropy increases the band depth and curvature of the magnon dispersion.
These adjustments to the magnon spectrum are expected to facilitate magnon condensate formation.
From the calculations of effective magnon-magnon interactions and minimizing the interaction potential at the band minima, we find the magnon condensate phase diagram. We demonstrate that the inclusion of PMA results in a magnon condensate with a more intricate phase diagram compared to when PMA is absent.
A finite PMA has the tendency to drive the magnon condensate towards a nonsymmetric magnon population at band minima in thinner films and lower magnetic fields, as compared to the absence of PMA.
Micromagnetic simulations within the LLG framework confirm our analytical results and analyses.

\section*{Acknowledgements}
The authors thank Anne Louise Kristoffersen for helpful discussions.
We acknowledge financial support from the Research Council of Norway through its Centers of Excellence funding scheme, project number 262633, "QuSpin". A.Q. was supported by the Norwegian Financial Mechanism Project No. 2019/34/H/ST/00515, "2Dtronics". P.P., A.A.S., and B.H. acknowledge financial support by
the Deutsche Forschungsgemeinschaft (DFG, German Research Foundation) TRR 173 Grant No. 268565370 Spin+X
(Projects B01 and B04).

\bibliography{refs_n}

\newpage
\appendix

\renewcommand{\thefigure}{S\arabic{figure}}
\setcounter{figure}{0}


\onecolumngrid
\section{Diagonalization of Magnon Hamiltonian}
The total spin Hamiltonian of a thin film, in the $y-z$ plane with a small perpendicular anisotropy along the $\uvec{x}$ direction reads,
\begin{equation}
\mathcal{H} = \mathcal{H}_\text{ex} + \mathcal{H}_\text{Z} + \mathcal{H}_\text{dip} + \mathcal{H}_\text{an}.
\end{equation}
The exchange energy between neighboring spins reads
\begin{equation}
\mathcal{H}_{\rm ex} = -\frac{1}{2} J_\text{ex} \sum_{i,j} \bm{S}_i \cdot \bm{S}_j ,
\end{equation}
where $ J_\text{ex} > 0$ is the ferromagnetic exchange constant and $a$ is the lattice constant.
The Zeeman energy due to an inplane external magnetic field of strength $H_0$ along the $\uvec{z}$ direction reads,
\begin{equation}
\mathcal{H}_\text{Z} = -g\mu_B H_0 \sum_j \bm{S}_j^z ,
\end{equation}
where $\mu_B$ is the Bohr magneton and $g$ is the the effective Land{\'e} g-factor.
The dipolar field is expressed as \cite{kreisel2009microscopic},
\begin{subequations}
\begin{align}
\mathcal{H}_{\rm dip} &=
-\frac{1}{2} \sum_{i,j} \sum_{\alpha,\beta}
D_{i,j}^{\alpha,\beta} S_i^\alpha S_j^\beta ,
\\
D_{i,j}^{\alpha,\beta} &= (g \mu_B)^2(1-\delta_{i,j})
\frac{\partial ^2}{\partial r_{ij}^\alpha \partial r_{ij}^\beta}
\frac{1}{\vert \bm{r}_{ij} \vert } ,
\end{align}
\end{subequations}
where $\alpha,\beta$ denote the spatial components $x, y $ and $z$;
and $\bm{r}_{ij}$ is the distance vector between the spin sites $i$ and $j$. Finally, the PMA anisotropy is given by,
\begin{equation}
\mathcal{H}_\text{an} = -K_\text{an} \sum_{j} (\bm{S}_j \cdot \hat{x})^2.
\end{equation}

The Holstein-Primakoff transformation allows us to express the spin operators in terms of bosonic creation and annihilation operators
$\op{a}^\dagger$ and $\op{a}$ respectively.
Using the large-$S$ approximation, we have,
$S^{+} \approx \hbar \sqrt{2S} (\op{a} - \op{a}^\dagger \op{a} \op{a}/(4S)  )$ ,
$S^{-} \approx \hbar \sqrt{2S} (\op{a}^\dagger - \op{a}^\dagger \op{a}^\dagger \op{a}/(4S)  )$, and
$S^{z} = \hbar(S - \op{a}^\dagger \op{a} )$ .

The corresponding noniteracting boson Hamiltonian in the Fourier space reads,
\begin{equation}
\mathcal{H}_2 = \sum_{\bm{q}}
A_{\bm{q}} \op{a}_{\bm{q}}^\dagger \op{a}_{\bm{q}}
+ \frac{1}{2} B_{\bm{q}} \op{a}_{\bm{q}} \op{a}_{\bm{-q}}
+ \frac{1}{2} B_{\bm{q}}^\ast \op{a}_{\bm{q}}^\dagger \op{a}_{\bm{-q}}^\dagger ,
\end{equation}
where
$\op{a}_j^\dagger = \frac{1}{N}\sum_{\bm{q}} e^{i \bm{k} \cdot \bm{r}_j}
 \op{a}_{\bm{q}}^\dagger$
and
 $A_{\bm{q}} $ ($B_{\bm{q}}$) is presented in Eq. (4) in the main text.
We utilize the following Bogoliubov transformation to diagonalize this bosonic Hamiltonian and find the corresponding noninteracting magnon Hamiltonian,
\begin{subequations}
\begin{align}
    \op{a}_{\bm{q}} = u_{\bm{q}} \op{c}_{\bm{q}} + v_{\bm{q}} \op{c}_{\bm{-q}}^\dagger ,\\
    \op{a}_{-\bm{q}}^\dagger = v_{\bm{q}}^\dagger \op{c}_{\bm{q}} + u_{\bm{q}} \op{c}^\dagger_{-\bm{q}},
\end{align}
\end{subequations}
where $u_{\bm{q}} = u_{\bm{-q}}$ and $v_{\bm{q}} = v_{\bm{-q}}$ are the Bogoliubov coefficients, given by,
\begin{subequations}
\begin{align}
u_{\bm{q}} &= \big(\frac{A_{\bm{q}} + 2\hbar \omega_{\bm{q}} }{2 \hbar \omega_{\bm{q}}} \big)^{\frac{1}{2}}, \\
v_{\bm{q}} &= \text{sgn}(B_{\bm{q}}) \big(\frac{A_{\bm{q}} - 2\hbar \omega_{\bm{q}} }{2 \hbar \omega_{\bm{q}}}\big)^{\frac{1}{2}}.
\end{align}
\end{subequations}
The Bogoliubov coefficients depend on the easy-axis magnetic anisotropy $K_{\rm{an}}$ via both magnon dispersion $\omega_{\bm{q}}$ and $A_{\bm{q}}$.
It can be shown that the off-diagonal terms ($\alpha \neq \beta$) of the dipolar interaction vanish in the uniform mode approximation for a thin film of infinite lateral lengths \cite{hick2010bose, kreisel2009microscopic}.
In this case, the dipolar interaction contains no three-magnon operator terms.
We omit any renormalization correction to the noninteracting magnon Hamiltonian, as they are of the order of $1/S$ and small.
We define the following parameters in the 4-magnon interaction amplitudes,
introduced in the main text Eq. (6), \cite{li2013phase, salman2017microscopic}.
\begin{subequations}
\begin{align}
\alpha_1 &= u_{\bm{Q}}^4 +  v_{\bm{Q}}^4 + 4 u_{\bm{Q}}^2  v_{\bm{Q}}^2, \\
\alpha_2 &= 2 u_{\bm{Q}}^2  v_{\bm{Q}}^2, \\
\alpha_3 &= 3 u_{\bm{Q}} v_{\bm{Q}} (u_{\bm{Q}}^2 +  v_{\bm{Q}}^2).
\end{align}
\end{subequations}
These parameters depend on $K_{\rm{an}}$ via Bogoliubov coefficients.

\section{The nonlinear interaction amplitudes}
\begin{figure}
\includegraphics[width=0.8\columnwidth]{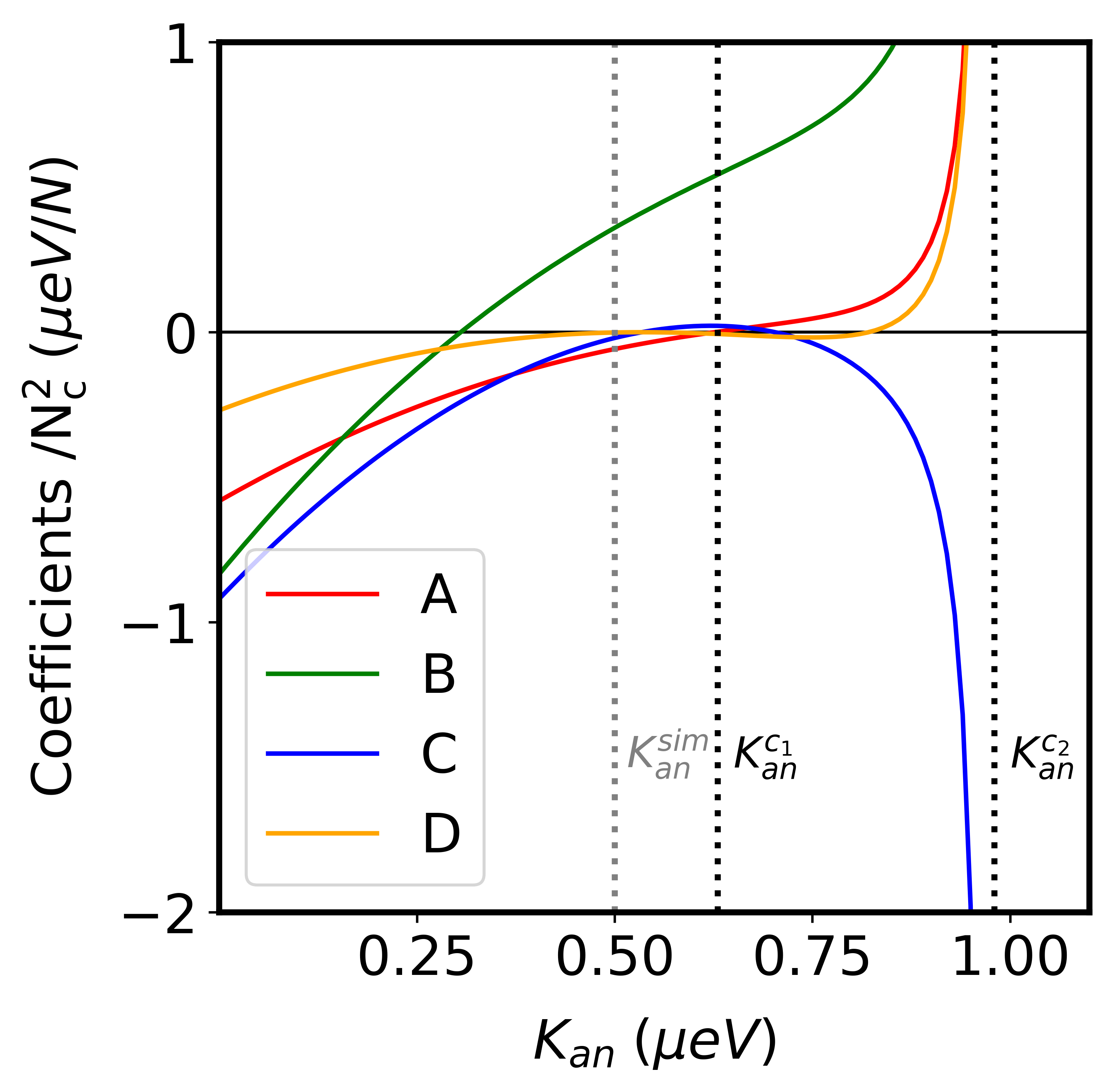}
\centering
\caption{
The interaction parameters, Eq. (6) in the main text, as a function of PMA. $A$ and $B$ are, respectively, the magnon-conserving intraband and interband interaction parameters while $C$ and $D$ are the magnon-non-conserving interband interaction parameters. We set $L_x = \SI{\thicknm}{\nano \meter}$ and $H_0 = \SI{1}{\kilo \Oe}$.
} \label{fig:plot1d_ABCD}
\end{figure}

In Fig. \ref{fig:plot1d_ABCD}, we plot the different intraband and interband interaction apmlitudes, see Eq. (6) in the main text, as a function of PMA.

\begin{table}
\begin{tabularx}{\columnwidth}{ll}
\hline \hline
Parameter & Value \\
\hline
Excitation time, first interval & $\SI{100}{\nano \second}$ \\
Excitation time, second interval & $\SI{200}{\nano \second}$ \\
Excitation strength, first interval & $-\SI{1.2e10}{\ampere \per \square \meter}$  \\
Excitation strength, second interval & $-\SI{0.35e10}{\ampere \per \square \meter}$ \\
\qquad $(K_{\rm an}=0)$ & ~ \\
Excitation strength, second interval & $-\SI{0.2e10}{\ampere \per \square \meter}$ \\
\qquad $(K_{\rm an}=\SI{0.5}{\micro \eV})$ & ~ \\
\hline \hline
\end{tabularx}
\caption{Simulation parameters} \label{tab:paramSim}
\end{table}

\section{Micromagnetic Simulations}
We solve the following LLG equation in the continuum model for the magnetization direction $\bm{m}(\bm{r},t)=\bm{S}/S$, using micromagnetic simulation code MuMax3 \cite{vansteenkiste2014design}, to study spin dynamics in the system,
\begin{align}
\frac{\partial\bm{m}}{\partial t}=\frac{\gamma}{1+\alpha^2} \big(\bm{m}\times \bm{B}_{\rm {eff}} + \alpha \bm{m}\times(\bm{m}\times \bm{B}_{\rm {eff}}) \big)+\bm{\tau}_{\rm {STT}},
\end{align}
where $\gamma$ is the gyromagnetic ration, $\alpha$ is the Gilbert damping parameter, $\bm{B}_{\rm {eff}}=- M_S^{-1} d \mathcal{H} / d \bm{m}$ is the effective magnetic field, $\bm{\tau}_{\rm {STT}}=\beta \bm{m} \times (\bm{m}_p\times \bm{m}) $ is  the Slonczewski spin-transfer torque, with $\beta$ is depends on material parameters and applied charge current density and $\bm{m}_p$ is the polarization of the spin current. In the continuum model, the exchange stiffness in the SI unit is related to the Heisenberg exchange interaction via $A_{\rm{ex}}=10^4 M_S D_{\rm{ex}}/(2\gamma)$.

We perform simulations of magnon creation by spin-transfer torque with and without out-of-plane anisotropy. We define an initial state in which the spins, on average, point along the $\uvec{z}$ direction. We introduce a random noise in the spin direction to mimic the thermal noise as the initial condition of our simulation. Next, we excite magnons in the ferromagnetic thin film by applying a spin torque, with $\bm{m}_p \| \uvec{z}$, to the entire film surface.
The film is discretized in the lateral directions and uniform in the $\uvec{x}$ direction. The lateral dimensions of the film are large compared to the film thickness. In this way, the film is effectively a 2D magnetic system.
The LLG equation is solved for each successive time step, using the open-source software MuMax3 \cite{vansteenkiste2014design}.
We refer to Ref. [\onlinecite{frostad2022spin}] for more numerical details.

The spin accumulation determines the strength of the spin torque, see Ref. [\onlinecite{frostad2022spin}]. We start the simulations by exciting magnons with a strong spin torque (interval $I_1$), before lowering the torque strength to keep the magnetization dynamics in a semi-stable state where the total number of magnons does not change dramatically over time (interval $I_2$).
In this semistable state, the two magnon populations may interact with each other, and the density of magnons in both minima may oscillate in time. However, the total number of magnons remains relatively unchanged. The current strength and time duration of the two intervals are listed in Table I in the SM. The magnetization data in Fig. 4 in the main text are from the last $\SI{50}{\nano \second}$ of the intervals.

The nonequilibrium magnon density in the film is proportional to the deviation of total magnetization for the ground state, $\eta = 1 - \langle m_z \rangle$ \cite{frostad2022spin}.

The torque strength determines the number of excited magnons. A finite PMA lowers the magnon spectrum minima, meaning that one can use a weaker spin torque to generate the critical magnon density needed for the generation of condensate magnons.
In Fig. 4 in the main text, we choose a spin-transfer torque strength for the generation of condensate magnons, resulting in a magnon density of approximately $\eta \approx \num{0.01}$. This density is small enough to only take into account two-magnon scattering possesses and not higher order processes.

The distribution of magnons in the magnon spectrum $\xi$ can be found by performing a Fourier transform of the transverse magnetization components,
$\xi(q_y,q_z;\omega) =
\vert \mathcal{F}[{m}_x(y,z;t)] \vert ^2 +
\vert \mathcal{F}[{m}_y(y,z;t)] \vert ^2 $ \cite{frostad2022spin}.

To reduce the consequences of the finite-size effect in our results, we analyze the magnetization data in the middle region of the thin film, $(y,z) \in [L/8, 7L/8]$, where $L=L_y=L_z=5 \rm{\mu m}$ are the lateral dimensions of the square thin film.

\end{document}